\documentclass{ltdPreprint_charlie}
\usepackage{graphicx}
\usepackage{float}
\usepackage{rotating}
\usepackage{amssymb}
\usepackage{mathptmx}
\usepackage{subcaption}
\usepackage{cite}
\begin{document}

\title{A large-diameter cryogenic rotation stage for half-wave plate polarization modulation on the POLARBEAR-2 experiment}

\author{C. A. Hill$^{a,b}$ \and A. Kusaka$^{b,c}$ \and P. Barton$^{b}$ \and B. Bixler$^{b}$ \and A. G. Droster$^{b}$ \and M. Flament$^{b}$ \and S. Ganjam$^{a,b}$ \and A. Jadbabaie$^{b}$ \and O. Jeong$^{a}$ \and A. T. Lee$^{a,b,d}$ \and A. Madurowicz$^{b}$ \and F. T. Matsuda$^{e}$ \and T. Matsumura$^{e}$ \and A. Rutkowski$^{b}$ \and Y. Sakurai$^{e}$ \and D. R. Sponseller$^{b}$ \and A. Suzuki$^{b}$ \and R. Tat$^{a,b}$}

\institute{$^{a}$Department of Physics, University of California, Berkeley, Berkeley, CA 94720, USA \\
\email{chill90@berkeley.edu} \\
$^{b}$Physics Division, Lawrence Berkeley National Laboratory, Berkeley, CA 94720, USA \\
\email{akusaka@lbl.gov} \\
$^{c}$Department of Physics, University of Tokyo, Bunkyo-ku, Tokyo 113-0033, Japan \\
$^{d}$Radio Astronomy Laboratory, University of California, Berkeley, Berkeley, CA 92093, USA \\
$^{e}$Kavli IPMU, University of Tokyo, Kashiwa, Chiba 277-8583, Japan}

\maketitle

\begin{abstract}

\vspace*{-3mm}
We describe the design of a cryogenic rotation stage (CRS) for use with the cryogenic half-wave plate (CHWP) polarization modulator on the POLARBEAR-2b and POLARBEAR-2c (PB2b/c) cosmic microwave background (CMB) experiments, the second and third installments of the Simons Array. Rapid modulation of the CMB polarization signal using a CHWP suppresses 1/f contamination due to atmospheric turbulence and allows a single polarimeter to measure both polarization states, mitigating systematic effects that arise when differencing orthogonal detectors. To modulate the full detector array while avoiding excess photon loading due to thermal emission, the CHWP must have a clear-aperture diameter of $> 450$ mm and be cooled to $< 100$ K. We have designed a $454$-mm-clear-aperture, $< 65$ K CRS using a superconducting magnetic bearing driven by a synchronous magnetic motor. We present the specifications for the CRS, its interfacing to the PB2b/c receiver cryostat, its performance in a stand-alone test, and plans for future work.


\end{abstract}

\vspace*{-3mm}
˚\section{Motivation}
\vspace*{-2mm}

Cosmic microwave background (CMB) polarization is a powerful probe of cosmology. Particularly, the ``B-mode'' CMB polarization pattern can be used to probe gravitational lensing on arcminute angular scales \cite{seljakGravPot, zaldProjMatDensity} and primordial gravitational waves on degree angular scales \cite{seljakGW, zaldarriagaAllSky, kamionkowskiCurl}. In order for a single telescope to characterize small and large scales simultaneously, it must observe with high resolution over a large sky area, requiring both a large primary aperture and good low-frequency (or ``1/f'') noise performance \cite{satoruHWP}.

Rapidly-rotating half-wave plates (HWP) are a common technique to modulate CMB polarization and reduce the impact of low-frequency noise on experiment sensitivity \cite{maxipolHWP, absHWP, ebexHWP, actHWP, satoruHWP, charlie, lbHWP}. A HWP up-converts the incident polarization signal into a white-noise-dominated frequency regime, hence suppressing 1/f contamination in the down-converted data \cite{absHWP}. Additionally, a HWP's birefringent substrate rotates the incident polarization vector, allowing a single polarimeter to measure both polarization states, therefore mitigating systematic effects that arise when differencing orthogonal detectors \cite{absHWP, absSyst, bryan}. HWPs have been adopted as continuous modulators on multiple large-aperture telescopes, including POLARBEAR \cite{satoruHWP, ziggy}, POLARBEAR-2a, \cite{charlie, yuki}, and the Atacama Cosmology Telescope \cite{kevinHWP, actHWP}.

However, if operated at ambient temperature, thermal emission from the HWP introduces parasitic photon loading onto the detector array, which degrades instrument noise performance \cite{charlie}. Therefore, cryogenic half-wave plates (CHWPs) are a key development for CMB polarization measurements \cite{ebexHWP, tomoHWP, bradHWP}. In this paper, we present a novel cryogenic rotation stage (CRS) for the CHWP on POLARBEAR-2b and POLARBEAR-2c (PB2b/c), the second and third installments of the Simons Array \cite{benSA, nate}. We discuss the CRS design, its interfacing to the PB2b/c receiver cryostat, its performance in a stand-alone test, and plans for future work.

\vspace*{-5mm}
\section{Requirements and Design}
\vspace*{-2mm}

The CRS, shown in Figure \ref{fig:RotStat}, uses a magnetic motor and a superconducting magnetic bearing (SMB) to drive rotation while reading out the rotor angle using an optical encoder. This system is non-contact and low-friction, minimizing heating while maximizing mechanical robustness. The requirements and design values for the CRS are driven by PB2b/c system performance and are shown in Table \ref{table:specs}.

\begin{table}[htbp]
\vspace*{-1mm}
\centering
	\begin{tabular}{|| c | c  | p{4.5cm} | c ||}
	\hline
	\bf{Parameter} & \bf{Requirement} & \multicolumn{1}{| c |}{\bf{Motivation}} & \bf{Design Value} \\
	\hline \hline
	Clear aperture diameter & $> 450$ mm & Negligible side lobes induced by beam truncation at the CHWP aperture & $454$ mm \\
	\hline
	Operating temperature & $< 100$ K & $< 1$ \% decrease in mapping speed due to CHWP thermal emission & $< 65$ K \\
	\hline
	Rotation frequency & $\geq 2$ Hz & Negligible 1/f contamination in the polarization modulation band & $\leq 3$ Hz \\
	\hline
	Continuous heating & $< 1$ W & $< 1$ K increase in the pulse tube cooler's first-stage temperature & $< 0.25$ W \\
	\hline
	Encoder timing jitter & $< 1 \: \mathrm{\mu s}$ & Noise induced by angle error $< 10$ \% of PB2b/c array-averaged noise equivalent CMB temperature & $0.2 \; \mathrm{\mu s}$ \\
	\hline
	\end{tabular}
	\vspace*{-2mm}
	\caption{The cryo-mechanical and readout requirements for the PB2b/c CRS. \label{table:specs}}
\vspace*{-11mm}
\end{table}

\iffalse
\begin{figure}[htbp]
\centering
\begin{subfigure}{.65\textwidth}
  	\centering
  	%\includegraphics[trim={1cm, 2cm, 0.5cm, 1cm}, clip, width=\linewidth]{RotorStator.pdf}
  	\includegraphics[trim={1cm, 5.5cm, 0.5cm, 4.5cm}, clip, width=\linewidth]{StandaloneCS.pdf}
   	\caption{\label{fig:RotStat}}
\end{subfigure}%
\begin{subfigure}{.35\textwidth}
  	\centering
  	\includegraphics[trim={5cm, 6cm, 5cm, 1cm}, clip, width=\linewidth]{Encoder.pdf}
  	\caption{\label{fig:Encoder}}
\end{subfigure}
\vspace*{-3mm}
\caption{Figure \ref{fig:RotStat} shows a computer-aided design (CAD) drawing of the CHWP rotator with key hardware elements highlighted. The tilt in the stator represents its buckling at 60 K due to the differential thermal contraction between the YBCO tiles and the aluminum fixture. Figure \ref{fig:Encoder} shows the optical encoder read head and highlights which photodiode-LED pairs are used for motor and angle encoding. The motor encoder waveform is generated by a lane of coarse-resolution slots in the slotted encoder plate shown in Figure \ref{fig:RotStat}, while the angle encoder waveform is generated by a lane of fine-resolution slots. Two encoder read heads are deployed for redundancy in case of an LED or photodiode failure. \label{fig:CRS}}
\vspace*{-5mm}
\end{figure}
\fi

\begin{figure}[htbp]
\centering
\begin{subfigure}{.50\textwidth}
  	\centering
  	\includegraphics[trim={1cm, 2cm, 0.5cm, 1cm}, clip, width=\linewidth]{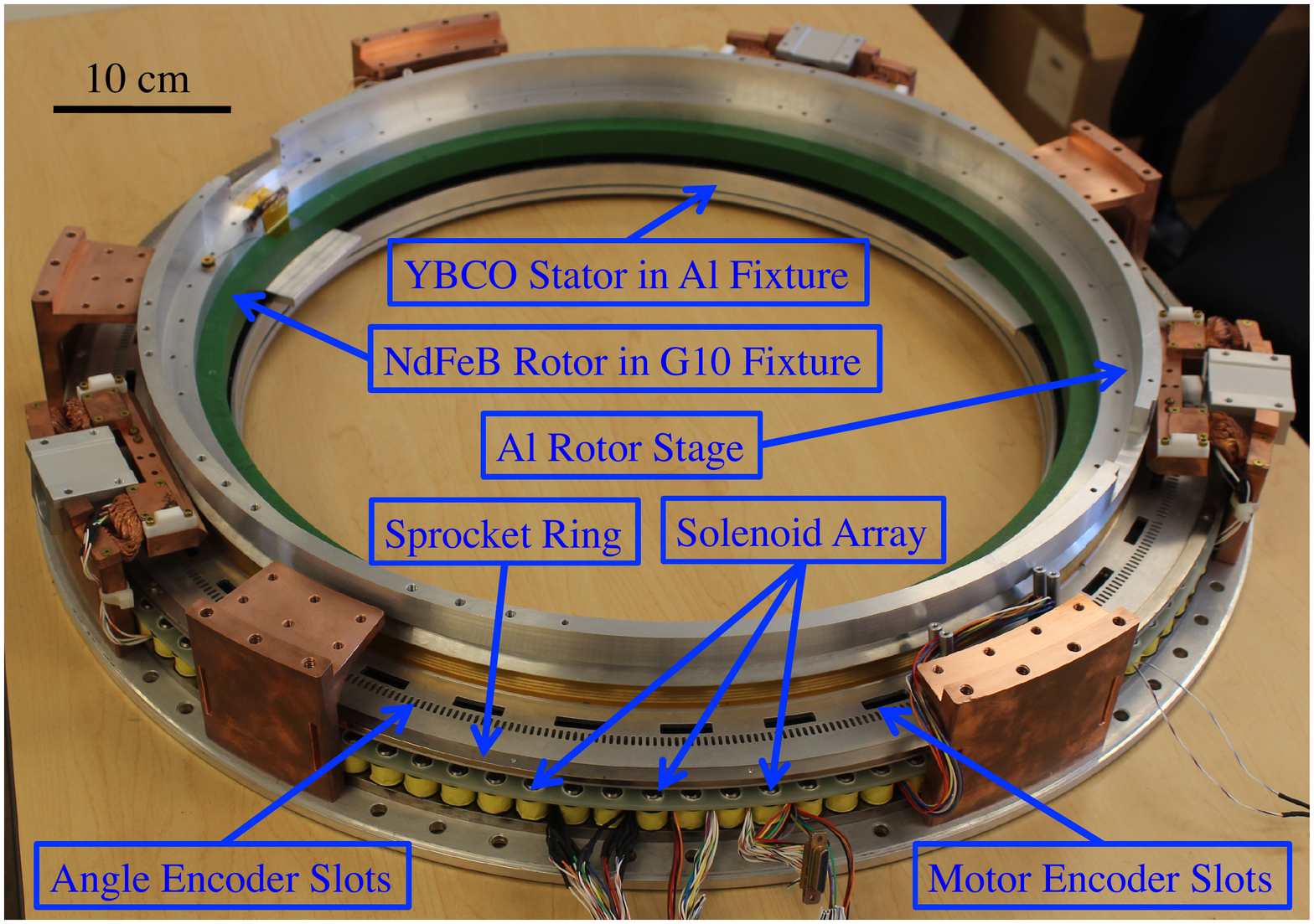}
   	\caption{\label{fig:RotStat}}
\end{subfigure}%
\begin{subfigure}{.50\textwidth}
  	\centering
  	\includegraphics[trim={5cm, 5cm, 5cm, 3.9cm}, clip, width=\linewidth]{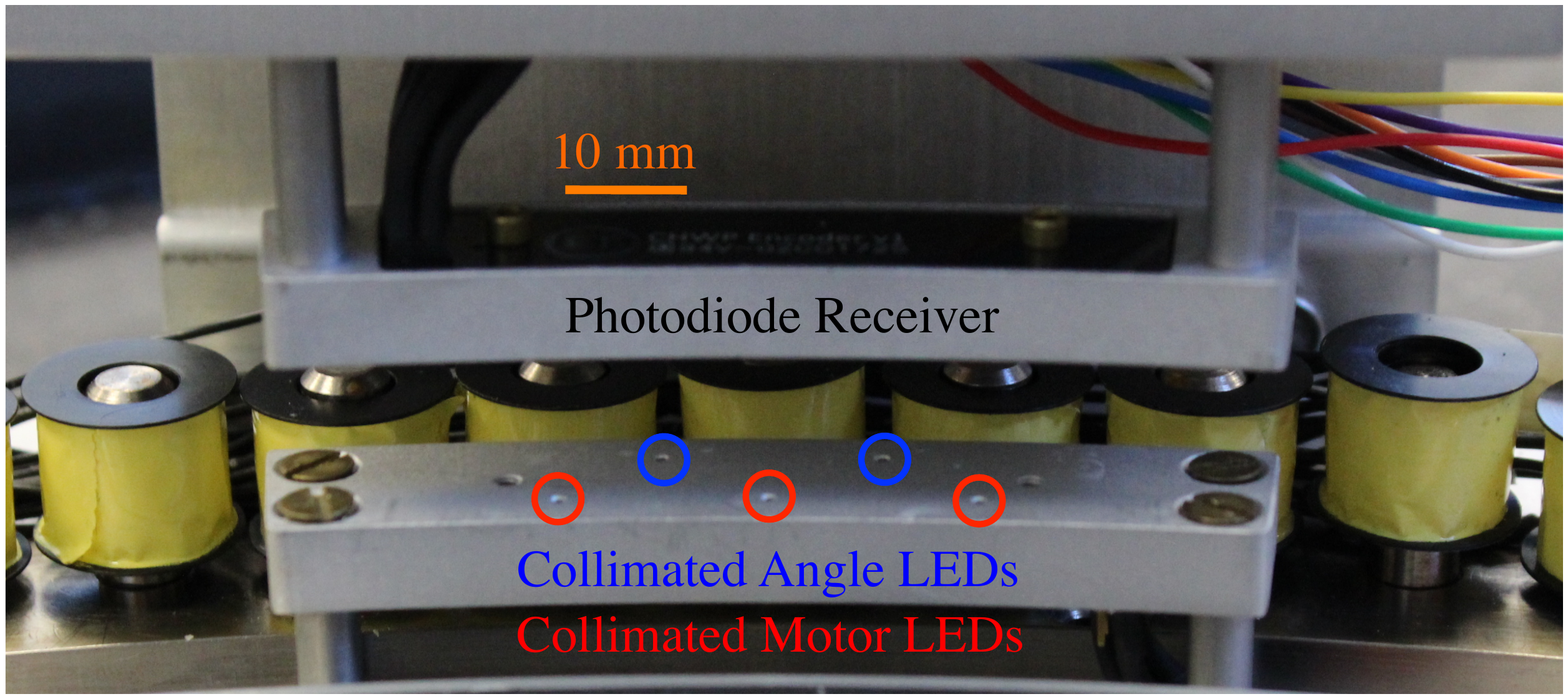}
  	\caption{\label{fig:Encoder}}
\end{subfigure}
\vspace*{-3mm}
\caption{Figure \ref{fig:RotStat} shows the CRS before insertion into the receiver cryostat with key hardware elements highlighted. The rotor is propped on three shims to maintain separation between the rotor and stator during assembly. Figure \ref{fig:Encoder} shows the optical encoder read head and highlights which photodiode-LED pairs are used for motor and angle encoding. The motor encoder waveform is generated by a lane of coarse-resolution slots on the rotor, while the angle encoder waveform is generated by a lane of fine-resolution slots. Two encoder read heads are deployed for redundancy in case of an LED or photodiode failure. \label{fig:CRS}}
\vspace*{-5mm}
\end{figure}

\vspace*{-2mm}
\subsection{Bearing}
\vspace*{-2mm}

The SMB operates by flux pinning in an azimuthally-symmetric geometry. By suspending a uniformly-magnetized ring (the rotor) above a normal type-II superconductor (the stator) and subsequently cooling the superconductor below its transition temperature, the ring's permanent magnetic field becomes frozen in the superconducting bulk, constraining the rotor in the axial and radial directions while allowing it to rotate freely in azimuth.

The SMB's effectiveness relies on small separation between the rotor and stator and on azimuthal homogeneity of magnetic flux in the superconductor. The SMB's spring constant scales with the distance between the magnet and superconductor as $1/z^{5}$, making small axial separation between the rotor and stator important to bearing stiffness \cite{hullFrozen}. Additionally, bearing friction scales with magnetic field inhomogeneity as $\Delta B_{perm}^{2}$ for eddy-current dissipation and $\Delta B_{perm}^{3}$ for hysteresis loss, making azimuthal homogeneity of the ring's magnetic field important to minimizing heating during rotation \cite{bean, zeisberger}.

The PB2b/c SMB is manufactured by Adelwitz Technologiezentrum GmbH \footnote{http://www.atz-gmbh.com/} (ATZ). The stator is a $566$-mm-outer-diameter (OD), $460$-mm-inner-diameter (ID), $15$-mm-deep ring consisting of $46$ tiles of yttrium barium copper oxide (YBCO) set in an aluminum fixture. The rotor is a $528$-mm-OD, $470$-mm-ID, $19$-mm-deep ring consisting of $16$ tiles of NdFeB set in a G10 fixture. The rotor is separated from the stator by $4$ mm, and the $20$-kg rotor deflects $\approx 1$ mm under gravity in the PB2b/c CRS assembly. The magnetic flux through the superconductor has an azimuthal root-mean-square homogeneity of $< 0.5\%$, resulting in small frictional heating, as discussed further in Section \ref{sec:meas}.

\vspace*{-2mm}
\subsection{Motor \label{sec:motor}}
\vspace*{-2mm}

\begin{figure}
\centering
	\includegraphics[trim={2cm, 4cm, 2cm, 5cm}, clip, width=0.95\linewidth]{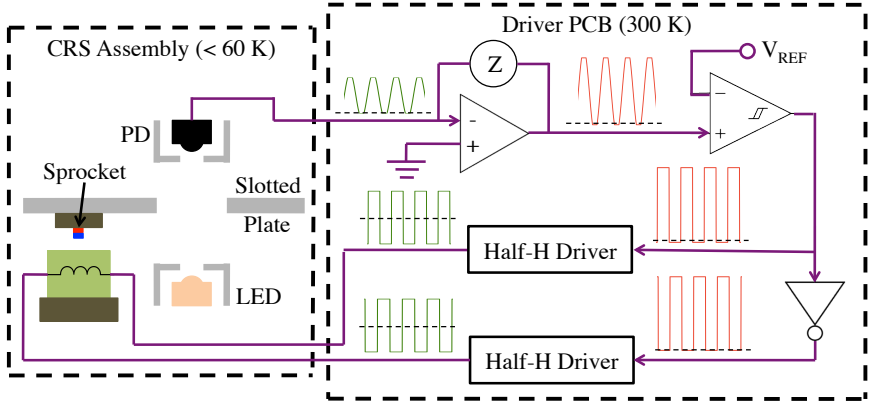}
	\vspace*{-8mm}
	\caption{A block diagram of one phase of the motor driver. A current-biased, collimated LED shines onto a collimated, reverse-biased photodiode (PD) though a lane of slots on the rotor. The photocurrent signal (green) is amplified to a voltage signal (red), is referenced to a comparator voltage $V_{\mathrm{REF}}$ to generate a TTL waveform, and is inverted. The inverted and non-inverted voltage signals are sent to two Half-H-bridge driver integrated circuits which are in series with the solenoid, creating a Full-H-bridge motor circuit. When the rotor rotates, the slots create an optically-chopped input to the photodiode, which the driver  uses to generate an alternating current in the solenoid. The AC excitation creates an alternating magnetic field that couples to the sprockets on the rotor and hence generates torque. \label{fig:motor}}
	\vspace*{-5mm}
\end{figure}

We operate an analog-feedback magnetic motor to drive the CRS, as shown in Figure \ref{fig:motor}. The active motor elements are 114 solenoids with low-carbon-steel cores equally spaced on a $634$-mm-diameter low-carbon-steel ring on the stator. The passive motor elements are 76 1.5-mm-diameter, 1.5-mm-tall NdFeB magnet ``sprockets,'' equally spaced with alternating polarity on a $634$-mm-diameter low-carbon-steel ring on the rotor. The solenoids are interpolated into three channels and are energized using three-phase alternating current. This scheme creates a rotating magnetic field on the stator that couples to the sprockets on the rotor, driving rotation.

The driver circuit uses waveforms from a cryogenic optical encoder to synthesize a Full-H-bridge alternating current to the solenoids. The motor encoder consists of three (one for each solenoid channel) collimated infrared (IR) light-emitting diodes (LEDs) shining onto three photodiodes (Figure \ref{fig:Encoder}) through a lane of coarse-resolution slots on the rotor (Figure \ref{fig:RotStat}). The read head design and photodiode circuitry used for the motor encoder is identical to that used for the angle encoder, which is described in Section \ref{sec:angenc}.

The motor encoder waveform is generated by $38$ slots (one for each pair of opposite-polarity sprockets) that create a $50\%$-duty-cycle, optically-chopped signal on the photodiode. The driver's phase is set by the relative position of the slots and sprockets and is chosen to maximize torque during startup. Because it is set by encoder feedback, the motor frequency is open-loop; therefore, rotation speed is set by tuning solenoid current amplitude. Given the measured torque due to friction described in Section \ref{sec:meas}, the driver current output is capable of spinning the rotor at $\leq 3$ Hz.

The upper-bound estimate of heat dissipation by the motor system onto the stator stage due to resistive losses in the coils during $2$ Hz continuous rotation at $60$ K is $< 100$ mW. Heat dissipated onto the levitating rotor is due primarily to eddy losses in the rotor's aluminum parts generated by the AC magnetic fields of the solenoids. This dissipation on the rotor is expected to be small, with an upper limit of $< 50$ mW.

\vspace*{-2mm}
\subsection{Angle Encoder \label{sec:angenc}}
\vspace*{-2mm}

The angle encoder consists of two collimated IR LEDs shining onto two photodiodes (Figure \ref{fig:Encoder}) through a lane of fine-resolution slots on the rotor (Figure \ref{fig:RotStat}). The encoder waveform is generated by $570$ slots that create a $50\%$-duty-cycle, optically-chopped, $1.1$ kHz signal on the photodiode when the rotor is spinning at $2$ Hz. One slit is omitted to determine absolute position, and the two photodiode-LED pairs are read in quadrature to ascertain rotation direction.

The LED emits at $940$ nm, and the photodiode is reverse-biased to linearize its response to incident flux intensity. The photodiode output is fed to a low-noise transimpedance amplifier in series with a comparator to synthesize a transistor-transistor logic (TTL) signal. This TTL waveform is sent to an Arduino Leonardo over Ethernet, which captures each rising and falling edge and timestamps them using its $16$ MHz internal clock. Simultaneously, the Arduino reads a Global-Positioning-System-synchronized, Inter-range instrumentation group (IRIG) timecode B waveform to record coordinated universal time. The Arduino collects 150 encoder packets before sending both the IRIG and angle data to an Intel Next Unit of Computing (NUC) miniature PC over Ethernet. The NUC interpolates the IRIG and angle data to reconstruct rotor orientation as a function of time.

The encoder readout is capable of identifying each slit edge to $0.2 \: \mathrm{\mu s}$. However, when demodulating the detector timestreams, the CHWP angle will need to be interpolated, as the $2.2$ kHz slit-edge frequency is slower than the $1$ MHz timing requirement. Therefore, rotational stability 
%on the timescale of PB2b/c's $\approx$ 5 ms detector time constant 
is critical to suppressing noise due to angle error in the demodulation routine.

\vspace*{-5mm}
\section{Interface to Receiver Cryostat}
\vspace*{-2mm}

\begin{figure}
\centering
	\includegraphics[trim={1.2cm, 2.4cm, 1.2cm, 2.9cm}, clip, width=0.95\linewidth]{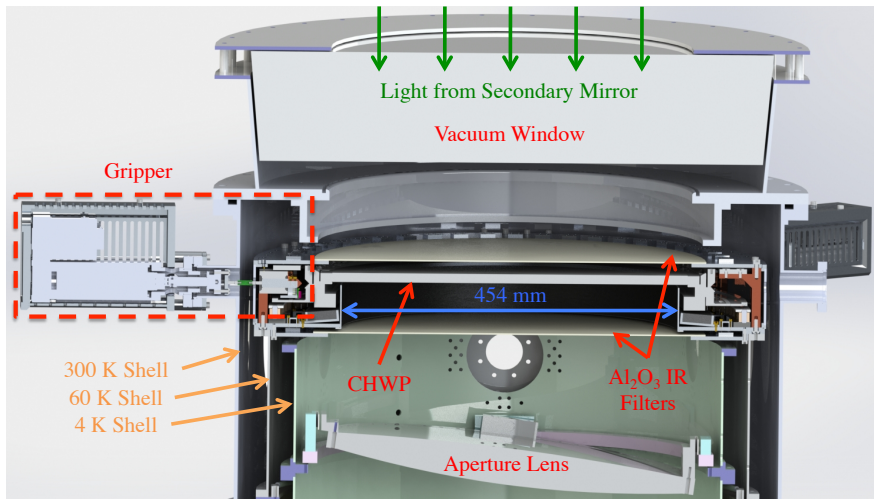}
	\vspace*{-8mm}
	\caption{The CHWP sits inside the PB2 cryostat window at the entry to the $60$ K shell, near the telescope secondary focus. It has one $< 60$ K alumina absorbing IR filter on either side of it to radiatively heat sink the levitating CHWP. When warm, the CHWP is aligned and held by a three-contact ``gripper'' system. The gripper consists of linear actuating vacuum feedthroughs, each coupled to a thermally-isolating rig, which in turn couples to a gripper ``finger'' in the $60$ K cavity. These fingers fit into an azimuthally-cut triangular groove on the rotor, mechanically constraining the rotor while providing thermal contact for cooling. The CHWP is un-gripped in this figure. \label{fig:pb2imp}}
	\vspace*{-5mm}
\end{figure}

The CRS is designed for use with the PB2b/c CHWP, which will be deployed in the PB2b/c receiver cryostat shown in Figure \ref{fig:pb2imp}. The CHWP is placed near the telescope secondary focus at the entry to the 60 K shell---whose temperature is $< 60$ K---between the 300 K cryostat window and the $4$ K aperture lens. $2$-mm-thick, $< 60$ K alumina IR filters on either side of the levitating CHWP create an IR-sealed cavity to which it is radiatively heat sunk. The optical cavity is defined by a $454$-mm-ID tube that hides rotating components from the focal plane. The CRS is designed to be compact along the optical axis to avoid vignetting the telescope's F/$1.9$ beam.

To both align the CHWP and secure it while warm, we implement a three-point ``gripper'' system, shown in Figure \ref{fig:pb2imp}. The gripper consists of three feedthrough linear actuators from Huntington Labs \footnote{https://huntvac.com/} sealed by a welded bellows. The feedthrough is coupled to a thermally-isolating rig that extends into the 60 K cavity, which in turn couples to a gripper ``finger.'' The finger is a precision-ground aluminum rod, which is connected to a copper wedge that is thermally sunk to 60 K via flexible oxygen-free high thermal conductivity copper braids. The aluminum rod is fed through a 60 K Frelon linear bearing, allowing the gripper finger to be actuated along the radial direction. When the CRS is warm, the gripper finger couples to an azimuthally-cut triangular groove on the rotor, securing and aligning the CHWP. As the 60 K stage cools, the gripper fingers provide thermal contact to the CHWP, and once the CHWP reaches base temperature, the gripper fingers are retracted, allowing the CHWP to rotate freely.

Maintaining the CHWP's base temperature during operation requires a stiff radiative coupling between the sapphire stack and the alumina IR filters. At $60$ K, the filters provide $\approx 20$ mW/K of cooling power to the rotor. Given the $< 50$ mW estimate for rotor heating described in Section \ref{sec:motor}, we expect the rotor to operate at a temperature of $< 65$ K, which meets the $< 100$ K requirement.

\vspace*{-5mm}
\section{System Characterization \label{sec:meas}}
\vspace*{-2mm}

\begin{figure}[htbp]
\begin{subfigure}{0.47\textwidth}
  	%\vspace*{\fill}
  	\centering
  	\includegraphics[trim={1cm, 1cm, 1cm, -1cm}, clip, width=\textwidth]{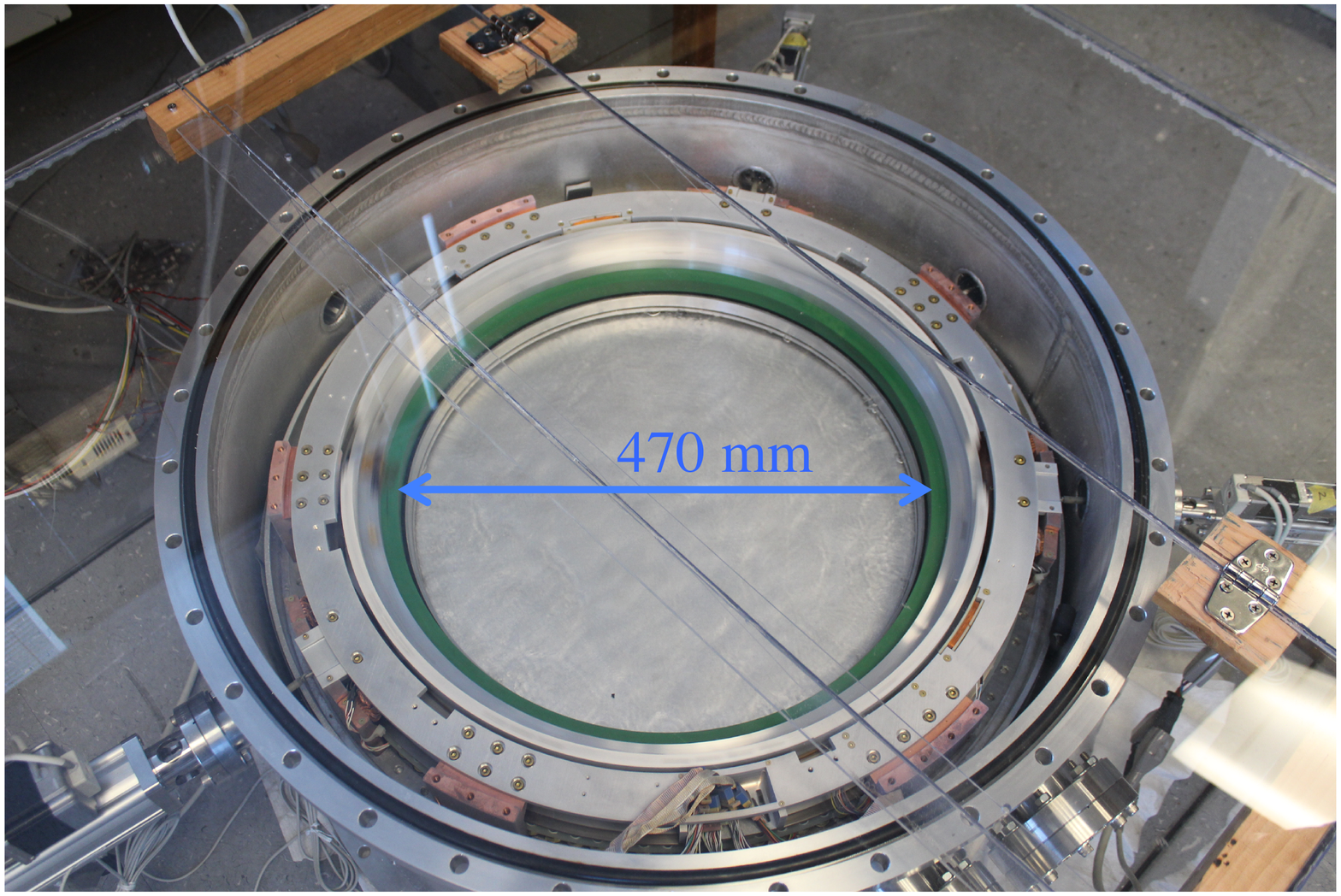}
  	\caption{\label{fig:teststand}}
	%\par\vfill
\end{subfigure}
\begin{subfigure}{0.53\textwidth}
	\centering
  	\includegraphics[trim={3.5cm, 3.2cm, 3.0cm, 2.5cm}, clip, width=\textwidth]{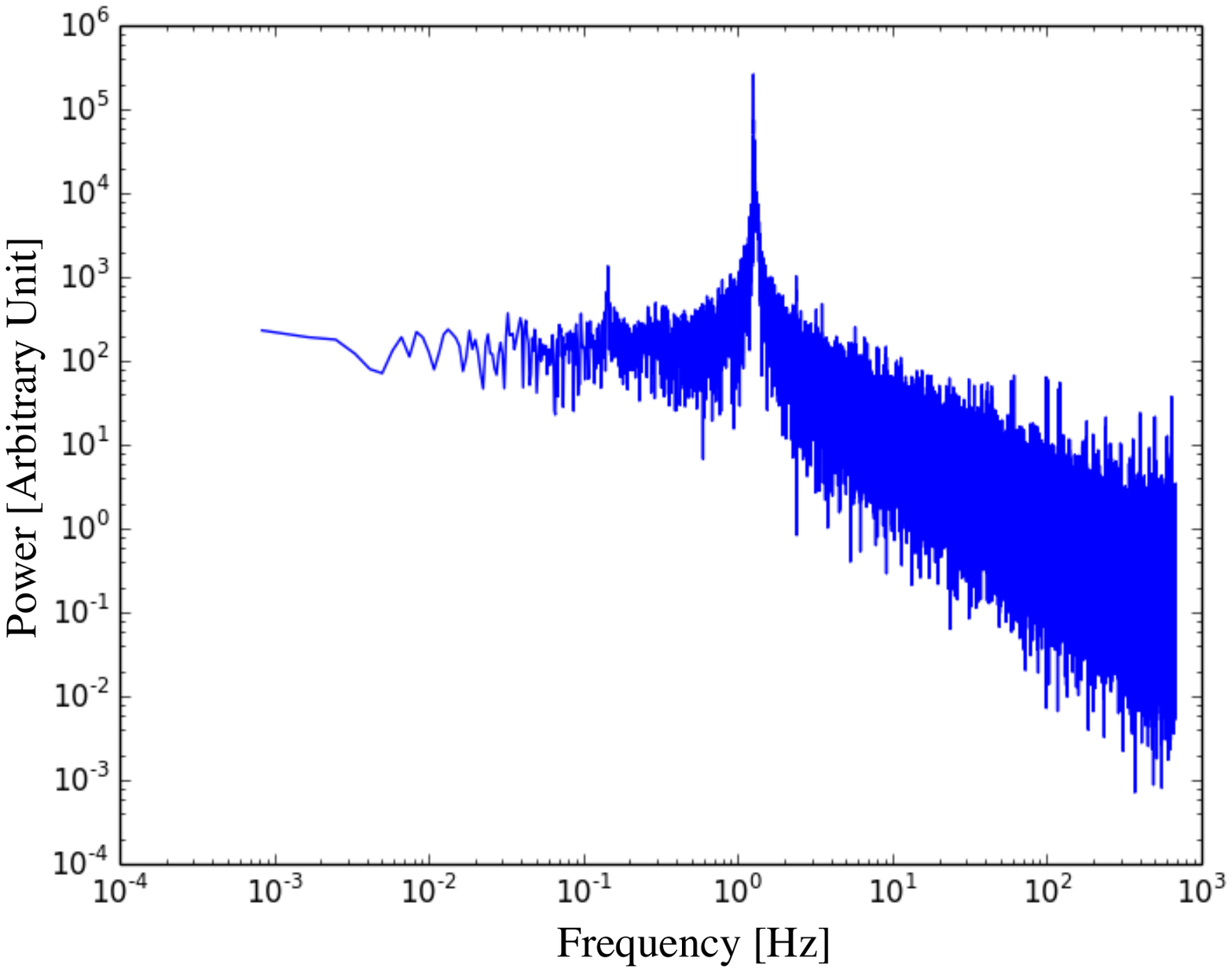}
  	\caption{\label{fig:controt}}
\end{subfigure}
\vspace*{-3mm}
\caption{Stand-alone testing using liquid nitrogen to cool the superconducting stator. Figure \ref{fig:teststand} shows the rotor spinning in the CRS setup. The assembly is enclosed in a polycarbonate ``dry box'' that traps evaporated nitrogen to prevent water condensation. Figure \ref{fig:controt} shows a power spectrum of angle encoder data taken while the rotor was spun at $\approx 1.2$ Hz continuously for $20$ minutes. The spectrum shows no readout pathologies and demonstrates wobble-free rotation and good low-frequency stability. \label{fig:testing}}
\vspace*{-5mm}
\end{figure}

Figure \ref{fig:teststand} shows an ambient stand-alone test setup used to validate key CRS functionalities prior to deployment in the PB2b/c receiver cryostat. The stator assembly is placed in a liquid nitrogen (LN2) bath, which is positioned within the 300 K shell using an adjustable-height stage. The entire assembly is enclosed in a polycarbonate box to create a dry, nitrogen-only environment at ambient pressure. This initial test is designed to demonstrate the operationality of the integrated system; a more precise evaluation of the thermal performance will require testing in a vacuum.

Testing involves gripping and aligning the rotor while warm, filling the LN2 bath to cool the stator, ungripping, ramping the motor, and continuously rotating while collecting encoder and IRIG data for $20$ minutes. The rotor was operated at $\approx 1.2$ Hz instead of 2 Hz due to the air resistance associated with operating at atmospheric pressure, as opposed to within a vacuum. A power spectrum of the encoder output during continuous rotation is shown in Figure \ref{fig:controt} and demonstrates no readout pathologies and good rotational stability. The results of this test establish the operationality of the integrated CRS system, its cryo-mechanical durability and gripper reliability, motor startup, continuous readout of the angle encoder, IRIG interfacing, and data acquisition quality.

\vspace*{-28mm}

Additionally, spin-down testing was performed to estimate friction as a function of rotation frequency in the rotor system. After subtracting a conservative estimate for torque due to air resistance, we find the frictional dissipation due to eddy currents and hysteresis loss to be $< 100$ mW at 2 Hz rotation. In total, our estimate for rotor friction, solenoid resistive heating, and eddy loss onto the rotor's aluminum parts add to $< 250$ mW at 2 Hz rotation, which is below the PB2b/c requirement of $1$ W. A more precise evaluation of the thermal performance of the CHWP system in a vacuum at the target rotation frequency of $2$ Hz is upcoming.

%\vspace*{-30mm}

\vspace*{-255mm}

\vspace*{-5mm}
\section{Conclusion}
\vspace*{-2mm}

\vspace*{-28mm}

We have designed, fabricated, and characterized key functionalities of a CRS for use with the PB2b/c CHWP. Design parameters meet the cryo-mechanical and readout requirements set by PB2b/c system performance. The CRS hardware has been integrated into a stand-alone test assembly, where the gripper, bearing, motor, and data-acquisition systems were validated as operational. Near-term developments for the CRS involve integration with the CHWP, deployment into the PB2b/c receiver cryostat, and characterization of mechanical heating, magnetic fields, and rotation-synchronous signals.

\vspace*{-30mm}

\begin{acknowledgements}
We gratefully acknowledge support for PB2b/c CRS development from the National Science Foundation through MSIP Grant \#1440338 as well as the Laboratory Directed Research and Development Program at Lawrence Berkeley National Laboratory.
\end{acknowledgements}

\pagebreak

\bibliographystyle{unsrt}
%\bibliography{CHWP_LTD_BIB_SHORT}

\begin{thebibliography}{99}

\bibitem{seljakGravPot}
{U. Seljak and M. Zaldarriaga}.
\newblock {Direct Signature of Evolving Gravitational Potential from Cosmic
  Microwave Background}.
\newblock {\em Phys.Rev.}, \textbf{D60}, 043504, (1999).

\bibitem{zaldProjMatDensity}
{M. Zaldarriaga and U. Seljak}.
\newblock {Reconstructing Projected Matter Density from Cosmic Microwave
  Background}.
\newblock {\em Phys.Rev.}, \textbf{D59}, 123507, (1999).

\bibitem{seljakGW}
{U. Seljak and M. Zaldarriaga}.
\newblock {Signature of Gravity Waves in Polarization of the Microwave
  Background}.
\newblock {\em Phys.Rev.Lett.}, \textbf{78}, 2054--2057, (1997).

\bibitem{zaldarriagaAllSky}
{M. Zaldarriaga and U. Seljak}.
\newblock {An All-Sky Analysis of Polarization in the Microwave Background}.
\newblock {\em Phys.Rev.}, \textbf{D55}, 1830--1840, (1997).

\bibitem{kamionkowskiCurl}
{M. Kamionkowski, A. Kosowsky, and A. Stebbins}.
\newblock {A Probe of Primordial Gravity Waves and Vorticity}.
\newblock {\em Phys.Rev.Lett.}, \textbf{78}, 2058--2061, (1997).

\bibitem{satoruHWP}
{S. Takakura et al.}
\newblock {Performance of a continuously rotating half-wave plate on the
  POLARBEAR telescope}.
\newblock {\em J.Cosmol.Astropart.Phys.}, \textbf{05}, 008, (2017).

\bibitem{maxipolHWP}
{B. R. Johnson et al.}
\newblock {MAXIPOL: Cosmic Microwave Background Polarimetry Using a Rotating
  Half-Wave Plate}.
\newblock {\em Astrophys.J.}, \textbf{665}, 42--54, (2007).

\bibitem{absHWP}
{A. Kusaka, T. Essinger-Hileman et al.}
\newblock {Modulation of CMB polarization with a warm rapidly-rotating
  half-wave plate on the Atacama B-Mode Search (ABS) instrument}.
\newblock {\em Rev. Sci. Instrum.}, \textbf{85}, 024501, (2014).

\bibitem{ebexHWP}
{K. MacDermid et al.}
\newblock {The performance of the bolometer array and readout system during the
  2012/2013 flight of the E and B experiment (EBEX)}.
\newblock {\em Proc. of SPIE}, \textbf{9153}, 915311, (2014).

\bibitem{actHWP}
{S. W. Henderson et al.}
\newblock {Advanced ACTPol Cryogenic Detector Arrays and Readout}.
\newblock {\em J. Low Temp. Phys.}, \textbf{184}, 772--779, (2015).

\bibitem{charlie}
{C. A. Hill, S. Beckman et al.}
\newblock {Design and development of an ambient-temperature
  continuously-rotating achromatic half-wave plate for CMB polarization
  modulation on the POLARBEAR-2 experiment}.
\newblock {\em Proc. of SPIE}, \textbf{9914}, 99142U-1, (2016).

%\bibitem{pancharatnam}
%{S. Pancharatnam}
%\newblock {Achromatic combinations of birefringent plates}
%\newblock {Memoir of the Raman Research Institute} \textbf{71}, 130--136, (1955).

\bibitem{lbHWP}
{T. Matsumura et al.}
\newblock {Design and performance of a prototype polarization modulator
  rotation system for use in space using a superconducting magnetic bearing}.
\newblock {\em IEEE Trans. Appl. Supercond.}, \textbf{26(3)}, (2016).

\bibitem{absSyst}
{T. Essinger-Hileman A. Kusaka et al.}
\newblock {Systematic effects from an ambient-temperature,
  continuously-rotating half-wave plate}.
\newblock {\em Rev. Sci. Instrum.,}, \textbf{87(094503)}, (2016).

\bibitem{bryan}
{S. Bryan et al.}
\newblock {A cryogenic rotation stage with a large clear aperture for the
  half-wave plates in the Spider Instrument}.
\newblock {\em Rev. Sci. Instrum.}, \textbf{87(014501)}, 10, (2016).

\bibitem{ziggy}
{Z. Kermish et al.}
\newblock {The POLARBEAR Experiment}.
\newblock {\em Proc. of SPIE}, \textbf{8452}, 84521C, (2012).

\bibitem{yuki}
{Y. Inoue et al.}
\newblock {POLARBEAR-2: an instrument for CMB polarization measurements}.
\newblock {\em Proc. of SPIE}, \textbf{9914}, 99141I, (2016).

\bibitem{kevinHWP}
{K. Coughlin and J. McMahon}.
\newblock {Metamaterial Acrhomatic Half-Wave Plates for Cosmic Microwave
  Background Observation}.
\newblock {\em Poster presented at: Low Temp. Det. 17}, \textbf{PD-1},  (2017).

\bibitem{tomoHWP}
{T. Matsumura et al.}
\newblock {Development of a half-wave plate based polarization modulator unit
  for LiteBIRD}.
\newblock {\em J. Low Temp. Phys.}, \textbf{This special issue}, (2017).

\bibitem{bradHWP}
{B. R. Johnson, F. Columbro et al.}
\newblock {A Large-Diameter Hollow-Shaft Cryogenic Motor Based on a
  Superconducting Magnetic Bearing for Millimeter-Wave Polarimetry}.
\newblock {\em Rev. Sci. Instrum.}, \textbf{88}, 105102, (2017).

\bibitem{benSA}
{B. Westbrook et al.}
\newblock {The POLARBEAR-2 and Simons Array Focal Plane Fabrication Status}.
\newblock {\em J. Low Temp. Phys.}, \textbf{This special issue}, (2017).

\bibitem{nate}
{N. Stebor et al.}
\newblock {The Simons Array CMB Polarization Experiment}.
\newblock {\em Proc. of SPIE}, \textbf{9914}, 99141H, (2016).

\bibitem{hullFrozen}
{J. R. Hull}.
\newblock {Effect of permanent-magnet irregularities in levitation force
  measurements}.
\newblock {\em Supercond. Sci. Technol.}, \textbf{13}, 854--856, (2000).

\bibitem{bean}
{C. P. Bean}.
\newblock {Magnetization of High-Field Superconductors}.
\newblock {\em Rev. Mod. Phys.}, \textbf{36(1)}, 31--40, (1964).

\bibitem{zeisberger}
{M. Zeisberger and W. Gawalek}.
\newblock {Losses in Magnetic Bearings}.
 \newblock {\em Mat. Sci. Eng.}, \textbf{53(1--2)}, 193--197, (1998).

\end{thebibliography}

\end{document}